# Higher-Order Harmonic Generation caused by Elliptically Polarized Electric Fields in Solid-State Materials

T. Tamaya,[1*] A. Ishikawa,[2] T. Ogawa,[3] and K. Tanaka[1,4]

[1] *Department of Physics, Graduate School of Science, Kyoto University, Sakyo-ku, Kyoto 606-8502, Japan*

[2] *Department of Science for Advanced Materials, University of Yamanashi, 4-3-11 Takeda, Kofu, Yamanashi 400-8511, Japan*

[3] *Department of Physics, Osaka University, Toyonaka, Osaka 560-0043, Japan*

[4] *Institute for Integrated Cell-Material Sciences, Kyoto University, Sakyo-ku, Kyoto 606-8501, Japan*

## ABSTRACT

We theoretically investigated the dependence of higher-order harmonic generation (HHG) in solid-state materials on the ellipticity of the electric field. We found that in the multiphoton absorption and ac Zener regimes, the intensity of HHG monotonically decreases with increasing ellipticity of the incident electric field, while in the semimetal regime, the intensity reaches a maximum for finite values of ellipticity. Moreover, the characteristics of the polarization of the emitted HHG change depending on the field intensity; only parallel emissions with respect to the major axis exist in the multiphoton absorption and ac Zener regimes, while both parallel and perpendicular emissions exist in the semimetal regime. These peculiar characteristics of the semimetal regime can be understood on the basis of changes in the HHG mechanism and may be able to be identified in experiments utilizing solid-state materials such as narrow-gap semiconductors.

One of the most fundamental and prominent aspects of nonlinear optics is higher-order harmonic generation (HHG) in gaseous media [1, 2]. The development of intense light sources has opened the way to studying non-perturbative optical phenomena through HHG [3-5] and has made it a cornerstone of various applications, such as attosecond pulse generation and molecular orbit tomography [6, 7]. In recent years, HHG in solid-state materials has been experimentally observed, and this has ushered in an era of strong-electric-field physics [8-14]. HHG in crystalline solids has been shown to have characteristics different from those of atomic matter, especially in terms of it reflecting the collective properties of the periodic arrangement of atoms. These differences have led to speculation on the generation mechanism and the possibility of controlling HHG [15-21]. Moreover, it is thought that a clear understanding of such matters is required for progress to be made on high-intensity optical technology.

A recent study had shown that the HHG mechanism in solid-state materials can be understood in terms of an analogy between Zener tunneling in semiconductors and tunnel ionization in gaseous media [22]. This consideration also indicates that high-intensity ac electric fields would introduce semimetallization of semiconductors corresponding to the over-the-barrier ionization processes [23, 24], which ensures the same characteristics of HHG as those of gaseous media. However, there is not enough experimental data on the HHG of over-the-barrier ionization processes because the high-intensity electric fields generate plasma and deplete neutral atoms, which break the phase-matching conditions [1, 2] and make it difficult to observe the characteristics of HHG. This difficulty could be avoided in experiments on solid-state materials if appropriate materials, such as narrow-gap semiconductors, are used. The unique characteristics of semimetallization of semiconductors should be emphasized by employing elliptically polarized electric fields, because the initial velocities of the excited carriers would vary depending on the electric field intensity and the trajectories predicted by the simple man model could be controlled by changing the ellipticity. These considerations indicate a necessity of investigating the dependence of HHG on the ellipticity of the incident electric field and point to the possibility of using HHG to explore unique characteristics of solid-state materials.

On the basis of the above considerations, in this paper, we theoretically investigate the dependence of HHG on the ellipticity of the incident electric field. Utilizing the theoretical framework developed in this paper, we find that HHG spectra arising from the semimetallization of semiconductors clearly show unique characteristics reflecting a change in the HHG mechanisms. The obtained predictions focus on peculiar aspects of HHG in solid-state materials that have never been seen in gaseous media.

To investigate the dependence of HHG on the ellipticity of the incident electric field, we will extend the theory employed in a previous work [22], which is only adequate for the case of linearly polarized light. For the derivation of the Hamiltonian, we start from the well-known formula $H = (1/2m_0)(\bm{p} - e\bm{A}(t)/c)^2 + \Sigma_i V(x - R_i)$, where $m_0$ is the electron mass, $e$ is the electron charge, $c$ is the velocity of light, $\bm{A}(t)$ is the vector potential of incident electric fields, $\boldsymbol{p}$ is the momentum of the bare electrons, and $V(x - R_i)$ is the periodic core potential of atoms located at $R_i$. In this framework, we will ignore the quasi-static energy $e^2\boldsymbol{A}(t)^2/2m_0c^2$ since it only shifts the total energy of the system [25]. Using the second quantization formulation, the Hamiltonian in the Coulomb gauge can be written as $H = H_0 + H_I$, where $H_0 = \int d\boldsymbol{x}\,\psi^\dagger(\boldsymbol{x})\{(1/2m_0)\boldsymbol{p}^2 + \Sigma_i V(\boldsymbol{x} - R_i)\}\psi(\boldsymbol{x})$ and $H_I = \int d\boldsymbol{x}\,\psi^\dagger(\boldsymbol{x})\{-(e/m_0c)\,\boldsymbol{A}(t)\cdot\boldsymbol{p}\}\psi(\boldsymbol{x})$. Here, $\psi(\boldsymbol{x})$ is the field operator of electrons. To consider the most basic structure of solid-state materials, we assume a two-dimensional covalent crystal in which two atoms A and B exist in a unit cell keeping the space-inversion symmetry. This assumption is equivalent to only focusing on the conduction and valence bands in a semiconductor. Using the tight-binding model only considering nearest-neighbor hopping of electrons and following the transformation $\hat{\psi}(\boldsymbol{x}) = (N)^{-1/2}\Sigma_{\boldsymbol{k},R_A}e^{i\boldsymbol{k}\cdot R_A}\phi(\boldsymbol{x} - R_A)\hat{a}_{\boldsymbol{k}} + (N)^{-1/2}\Sigma_{\boldsymbol{k},R_B}e^{i\boldsymbol{k}\cdot R_B}\phi(\boldsymbol{x} - R_B)\hat{b}_{\boldsymbol{k}}$, where the wave functions of the electrons bound to atoms A and B are described as $\phi(\boldsymbol{x} - R_A)$ and $\phi(\boldsymbol{x} - R_B)$, we can arrive at the tight-binding Hamiltonian,

$$H_0 = \Sigma_{\boldsymbol{k}}\left(\gamma f(\boldsymbol{k})a_{\boldsymbol{k}}^\dagger b_{\boldsymbol{k}} + \gamma f^*(\boldsymbol{k})b_{\boldsymbol{k}}^\dagger a_{\boldsymbol{k}}\right), \tag{1}$$

$$H_I = -\hbar\Sigma_{\boldsymbol{k}}\left(\Omega_R(\boldsymbol{k},t)a_{\boldsymbol{k}}^\dagger b_{\boldsymbol{k}} + h.c.\right). \tag{2}$$

Here, $\gamma$ is the transfer integral, $\hbar$ is the Planck constant, $f(\boldsymbol{k})$ is a form factor defined as $f(\boldsymbol{k}) = \Sigma_i e^{i\boldsymbol{k}\cdot\boldsymbol{\delta}_i} = |f(\boldsymbol{k})|e^{i\theta_{f(\boldsymbol{k})}}$, $\boldsymbol{\delta}_i$ is a lattice vector, $a_{\boldsymbol{k}}$ ($b_{\boldsymbol{k}}$) is the annihilation operator of electrons with wavenumber $\boldsymbol{k}$ on the sub lattice A (B), and $\Omega_R(\boldsymbol{k},t)$ is the Rabi frequency [26] defined by $\Omega_R(\boldsymbol{k},t) = (e\hbar\pi(\boldsymbol{k},t)/m_0c)$, where $\pi(\boldsymbol{k},t) = \Sigma_i e^{i\boldsymbol{k}\cdot\boldsymbol{\delta}_i}\int d^2x\,\phi^*(\boldsymbol{x})\boldsymbol{A}(t)\cdot\boldsymbol{p}\phi(\boldsymbol{x}-\boldsymbol{\delta}_i)$. In the following formulation, we will ignore the $\boldsymbol{k}$ dependence of the Rabi frequency, which is usually permitted in semiconductor physics [27]. Supposing the vector potential of the elliptically polarized light to be $\boldsymbol{A}(t) = A_0(1 + \epsilon^2)^{-1/2}\exp(-(t - t_0)^2/\tau^2)(\cos\omega_0 t, \epsilon\sin\omega_0 t)$, we can arrive at the Rabi frequency in the form, $\Omega_R(t) = \Omega_{R0}(t)(1 + \epsilon^2)^{-1/2}(\cos\omega_0 t + i\epsilon\sin\omega_0 t)$, where $\Omega_{R0}(t) = \Omega_{R0}\exp(-(t - t_0)^2/\tau^2)$. Throughout this paper, the parameters of the incident electric field are fixed to $t_0 = 12\pi/\omega_0$ and $\tau = 4\pi/\omega_0$. Here, we assume the major axis to be the *x*-axis and denote the ellipticity of the incident electric field as $\epsilon$ varying in the range of $-1 \leq \epsilon \leq 1$. The plus and minus signs of the ellipticity indicate left- and right-handed elliptically polarized light; therefore, the same dependence would be expected under space-inversion symmetry in solid-state materials. The transformation from the lattice picture to the band-structure picture can be performed by diagonalization of the single-particle part $H_0$ through the use of the electron-hole picture defined as $e_k = 1/\sqrt{2}[a_k + e^{i\theta_{f(k)}} b_k]$ and $h^\dagger_{-k} = 1/\sqrt{2}[-a_k + e^{i\theta_{f(k)}} b_k]$, where $e_k(h_k)$ is the annihilation operator of electrons (holes) with the Bloch wavevector $\boldsymbol{k}$. To consider the most simplified case, we assume $\gamma|f(\boldsymbol{k})| \approx (\hbar^2 k^2/2m_\mu + E_g/2)$ and $\theta_{f(k)} = \theta_k$. On the basis of this transformation, we can derive a Hamiltonian in the form of $H = H_0 + H_I$, where

$$H_0 = \Sigma_k \left((E^e_k + E_g/2)\, e^\dagger_k e_k + (E^h_k + E_g/2)\, h^\dagger_{-k} h_{-k}\right), \qquad (3)$$

$$\begin{aligned}H_I &= \hbar\Omega_{R0}(t)\Sigma_k (1+\epsilon^2)^{-1/2}(\cos\omega_0 t \cos\theta_k + \epsilon \sin\omega_0 t \sin\theta_k)(e^\dagger_k e_k + h^\dagger_{-k} h_{-k} - 1) \\ &\quad + i\hbar\Omega_{R0}(t)\Sigma_k (1+\epsilon^2)^{-1/2}(\cos\omega_0 t \sin\theta_k - \epsilon \sin\omega_0 t \cos\theta_k)(e^\dagger_k h^\dagger_{-k} - h_{-k} e_k).\end{aligned} \qquad (4)$$

Here, $E^\sigma_k = \hbar^2 k^2/2m_\sigma (\sigma = e, h)$ are the kinetic energies of electrons and holes with Bloch wavevector $\boldsymbol{k}$ and $E_g$ is the band-gap energy. The zero value of ellipticity can recover the Hamiltonian in Ref. [22]. The first term on the right-hand side in Eq. (4) indicates the intraband transition caused by elliptically polarized light, which can be renormalized in terms of single-particle energies $\epsilon^\sigma_k$ in the form, $\epsilon^\sigma_k(t) = E^\sigma_k + E_g/2 + \hbar\Omega_{R0}(t)(1+\epsilon^2)^{-1/2}(\cos\omega_0 t \cos\theta_k + \epsilon \sin\omega_0 t \sin\theta_k)$. This expression corresponds to the temporal variations of the band structures reflecting the features of the incident electric fields, wherein linearly polarized light ($\epsilon = 0$) causes variations only along the direction of the major axis ($k_x$-axis), while circularly polarized light ($\epsilon = 1$) causes anisotropic deformations in the $\boldsymbol{k}$ plane. These variations can be ignored when focusing on a weak electric field where the Rabi frequency is much smaller than the band-gap energy, i.e., $\hbar\Omega_{R0}(t) \ll E_g$. In this case, the single-particle energies are transformed into $\epsilon^\sigma_k(t) \approx E^\sigma_k + E_g/2$, and therefore, the optical and semiconductor Bloch equations [26] can be recovered if the factors $\sin\theta_k$ and $\cos\theta_k$ are ignored. This consideration indicates that the optical and semiconductor Bloch equations are only adequate for weak electric fields and would be violated by strong electric fields.

Using the above Hamiltonian, the time evolution equations of the densities $f^\sigma_k = \langle \sigma^\dagger_k \sigma_k \rangle$ and polarization $P_k = \langle h^\dagger_{-k} e_k \rangle$ with a Bloch wavevector $\boldsymbol{k}$ can be derived as

$$i\frac{\partial}{\partial t}P_k = [\epsilon^e_k(t) + \epsilon^h_k(t)]P_k + i\Omega_{R0}(t)(1+\epsilon^2)^{-1/2}(\cos\omega_0 t \sin\theta_k - \epsilon \sin\omega_0 t \cos\theta_k) \\ \times [1 - f^e_k - f^h_k] - i\gamma_t P_k, \qquad (5)$$

$$\frac{\partial}{\partial t}f^\sigma_k = 2\text{Im}[i\Omega_{R0}(t)(1+\epsilon^2)^{-1/2}(\cos\omega_0 t \sin\theta_k - \epsilon \sin\omega_0 t \cos\theta_k)P^\dagger_k] - \gamma_l f^\sigma_k. \qquad (6)$$

Here, $\gamma_t$ and $\gamma_l$ are the transverse and longitudinal relaxation constants, and in this study, they are fixed to $\gamma_t = 0.1\omega_0$ and $\gamma_l = 0.01\omega_0$. The numerical solutions of these equations give the time evolutions of the distributions of the carrier densities and polarization in two-dimensional $\boldsymbol{k}$ space. Figures 1(a1) and 1(b1) show the distributions of densities $f_k^e = f_k^h$, and Figs. 1(a2) and 1(b2) show the distributions of polarizations $2\text{Im}[P_k]$ in the cases of linearly and circularly polarized light for $t = 12\pi/\omega_0$. Here, we have set the Rabi frequency and the band-gap energy to $\Omega_{R0} = 2\omega_0$ and $E_g = 5\omega_0$. These figures indicate that for the case of linearly polarized light, the anisotropic distributions of densities and polarization are only linked to the direction of the polarization axes ($k_x$-axis), while for the case of circularly polarized light, the distributions resemble a spiral reflecting the temporal evolutions of the elliptically polarized light. These anisotropies are caused by dipole transitions under the temporal variations of the band structures characterized by the single-particle energies as $\epsilon_k^\sigma(t) = E_k^\sigma + E_g/2 + \hbar\Omega_{R0}(t)(1+\epsilon^2)^{-1/2}(\cos\omega_0 t\cos\theta_k + \epsilon\sin\omega_0 t\sin\theta_k)$. Utilizing these distributions, the time evolution of the generated currents along the $x$- and $y$- axes can be calculated using the definition, $J_\nu(t) = -c\langle\partial H_I/\partial A_\nu\rangle (\nu = x, y)$, i.e., $J_x(t) \propto \sum_k[(1 - f_k^e - f_k^h)\cos\theta_k - 2\text{Im}(P_k)\sin\theta_k]$ and $J_y(t) \propto \epsilon\sum_k[(1 - f_k^e - f_k^h)\sin\theta_k + 2\text{Im}(P_k)\cos\theta_k]$. The numerical results in Figs. 1(a3) and (b3) are for linearly and circularly polarized light, respectively. These figures indicate that currents are only generated along the major axis in the case of the linearly polarized light, while in the case of elliptically polarized light, they occur along both the $x$- and $y$- axes and exactly reflect the phase differences of the incident electric fields characterized by $E_0(\cos\omega_0 t, \sin\omega_0 t)$. Thus, we can obtain the time evolutions of the current in the form, $J(t) = (J_x(t), J_y(t))$, and can calculate higher-order harmonic spectra from the definition, $I(\omega) = \omega^2|J(\omega)|^2$, where $J(\omega) = (J_x(\omega), J_y(\omega))$ is the Fourier transform of the current vector $J(t)$. As indicated in a previous paper [22], the HHG mechanism can be classified into three regimes depending on the Rabi frequency: (i) the multiphoton absorption regime ($\Omega_{R0} \leq 0.5\omega_0$), (ii) ac Zener regime ($0.5\omega_0 \leq \Omega_{R0} \leq E_g/2\hbar$), and (iii) semimetal regime ($E_g/2\hbar \leq \Omega_{R0}$), from which we can expect the dependence of HHG on the ellipticity to vary according to the field intensity. Below, we discuss the characteristics of the higher-order harmonic spectra in each regime, setting the band-gap energy to $E_g = 10\omega_0$.

Figure 2(a) shows the higher-order harmonic spectra in the multiphoton absorption regime for ellipticity values of $\epsilon = 0$ (red line), $\epsilon = 0.4$ (green line), and $\epsilon = 0.8$ (blue line) in the case of $\Omega_{R0} = 0.4\omega_0$. This figure clearly shows that the intensities of the $N$th-ordered harmonics monotonically decrease with increasing ellipticity. Moreover, the decreasing ratio becomes larger as the harmonic order increases. To investigate the decreasing ratio in more detail, we plot in Fig. 3(a) the ellipticity dependence of the 7$^{th}$-order harmonics (blue line), which can be divided into parallel (red line) and perpendicular (green line) emissions with respect to the major axis (*x*-axis). This figure shows a monotonic decrease in total HHG intensity with increasing ellipticity as well as only parallel emissions. These tendencies can be understood on the basis of perturbative theory in gaseous media, which leads to the following expression for the *n*th-order polarization: $P^{(n)}(n\omega) = (\boldsymbol{e}_x - i\epsilon\,\boldsymbol{e}_y)(1 + \epsilon^2)^{-1/2}E_0^{(n-1)}[(1-\epsilon^2/1+\epsilon^2)]^{(n-1)/2}\chi^{(n)}(-n\omega)$, where $\chi^{(n)}(n\omega)$ and $\boldsymbol{e}_\nu$ are the *n*th ordered susceptibility caused by linearly polarized electric fields and a unit vector in the direction of the $\nu$ axis [28]. This indicates an ellipticity dependence of the HHG intensity of the form, $I_N^{total} \propto [((1-\epsilon^2)/(1+\epsilon^2))]^{N-1}$ and emission ratio between parallel and perpendicular components of $I_{Ny}/I_{Nx} \propto |P_y^N|^2/|P_x^N|^2 = \epsilon^2$. These expressions can explain the characteristics of the HHG in our numerical calculations. Note that they also indicate that circularly polarized electric fields, with $\epsilon = 1$, completely suppress HHG because of the condition $P^{(n)}(n\omega) = 0$. This characteristic can be also identified in our numerical calculations.

Figure 2(b) shows the higher-order harmonic spectra in the ac Zener regime, where the ellipticity values are $\epsilon = 0$ (red line), $\epsilon = 0.4$ (green line), and $\epsilon = 0.8$ (blue line) in the case of $\Omega_{R0} = 2\omega_0$. This figure clearly indicates the cutoff energy and the intensities of *n*th-order HHG decrease with increasing ellipticity. We also plot in Fig. 3(b) the ellipticity dependence of the 13th higher-order harmonic (blue line), which can be divided into parallel (red line) and perpendicular (green line) emissions with respect to the major axis. Here, we chose the 13th harmonic for its cutoff energy, evaluated as $E_c \approx E_g + 1.6\hbar\Omega_{R0}$ [22]. This figure shows a monotonic decrease in the total intensity of HHG, whose shape is roughly Gaussian and has already been identified in a recent experiment [14]. Moreover, we can see there are only parallel emissions with respect to the major axis. These characteristics are very similar to those of gaseous media [29-34]; thus, the semiclassical picture of the simple man model [35] seems to be useful for explaining the tendencies of HHG. As indicated by several researchers [29-34], the simple man model shows that the origin of HHG is recombination processes of accelerated carriers, and small ellipticity gives large displacements from the recombination positions, as is shown in Fig. 4(a). Accordingly, we can confirm that the cutoff energy and the intensities of HHG become strongly suppressed with increasing ellipticity and their suppression ratio would have a Gaussian shape [34], which can be seen in Fig. 2(b) and Fig. 3(b). The ellipticity dependence of the 13th harmonics being mainly due to parallel emission and almost no perpendicular emission supports this conjecture because the trajectories of the carriers would be closed only when they are driven by linearly polarized electric fields. Note that our numerical calculations also show that a circularly polarized electric field completely suppresses HHG. This tendency is the same as in the multiphoton absorption regime and has already been observed in an experiment utilizing a ZnO crystal wherein a luminescence peak remained around the band-gap energy [8].

The higher-order harmonic spectra in the semimetal regime are plotted in Fig. 2(c) in the case of $\Omega_{R0} = 8\omega_0$ for ellipticity values of $\epsilon = 0$ (red line), $\epsilon = 0.3$ (green line), and $\epsilon = 0.6$ (blue line). Different from the multiphoton absorption and ac Zener regimes, the spectra show a non-monotonic tendency with increasing ellipticity, where their intensities appear to be enhanced around $\epsilon = 0.3$. To investigate this tendency in detail, we plot in Fig. 3(c) the ellipticity dependence of the 28th harmonics (blue line), which can be divided into parallel (red line) and perpendicular (green line) emissions with respect to the major axis. The 28th order was evaluated from the cutoff energy, estimated in the form $E_C \approx 3.6\hbar\Omega_{R0}$ [22]. The ellipticity dependence of the total HHG intensities clearly reaches a maximum around $\epsilon = 0$ and $\epsilon = 0.3$. Moreover, both perpendicular and parallel emissions can be identified, especially for $\epsilon = 0.3$. These tendencies have never been observed in the multiphoton absorption and ac Zener regimes, while a few experiments performed in gaseous media [31, 32] have reported similar characteristics and suggested a variation in the HHG mechanisms between the lower and higher order harmonics. Spurred by this consideration, we plot in Fig. 5(a) the harmonic order dependence of HHG (blue line), which can be divided into parallel (red line) and perpendicular (green line) emissions in the case of $\epsilon = 0.3$. This figure shows that the lower order harmonics mainly derive from parallel emissions, while the higher order harmonics derive only from perpendicular emissions. This tendency is clearly seen in the ellipticity dependences of the 5th, 15th, and 30th harmonics plotted in Figs. 5(b), 5(c), and 5(d).

The physical interpretation of these tendencies would be on the basis of changes in the HHG mechanism. In the semimetal regime, the carriers are excited when the band gap closes, which indicates a long interval of overlap between the conduction and valence bands. This interval has never been seen before in the ac Zener regime, where excitations mainly occur when the *x*- and *y*-components of the incident electric field reach a maximum and zero, respectively. Here, we suppose the condition $\epsilon \ll 1$. In this case, the initial velocity of the excited carriers only has parallel components with respect to the major axis. On the other hand, in the semimetal regime, the long interval of overlap between the conduction and valence bands ensures a finite time period for the excitation processes and enhances both the parallel and perpendicular components of the initial velocity. As shown in Ref. [29-34], the perpendicular component of the initial velocity requires the ellipticity to be finite for the closed trajectories and elliptical orbits in real space (Fig. 4(b)). On the basis of this consideration, we can conclude that HHG in the semimetal regime is enhanced by finite ellipticity, while the HHG in the ac Zener regime monotonically decreases with increasing ellipticity. We expect that the higher harmonics near the cutoff energy mainly derive from the perpendicular components which do not appear in the ac Zener regime, while the lower harmonics derive mainly from the parallel components. Note that the numerical calculations do not show that circularly polarized light completely suppresses HHG.

In conclusion, we theoretically investigated the dependence of HHG in solid-state materials on the ellipticity of the incident electric field. By utilizing the framework developed in this paper, we found that the dependence of HHG intensity changes according to the ellipticity of the electric field. In the multiphoton absorption and ac Zener regimes, the HHG intensity monotonically decreases with increasing ellipticity and only emissions parallel to the major axis exist. On the other hand, in the semimetal regime, the intensity exhibits pronounced maximum peaks for finite values of ellipticity and both parallel and perpendicular emissions exist. The differences in these characteristics can be understood on the basis of changes in the HHG mechanisms of solid-state materials. The unique characteristics obtained here in the semimetal regime, which would be difficult to observe in gaseous media, should be able to be observed in narrow- and zero-gap semiconductors, such as InSb and graphene.

**ACKNOWLEDGMENTS**

This work was supported by a Grant-in-Aid for Scientific Research (A) (Grant Nos. 26247052 and 23244065) and a Grant-in-Aid for Scientific Research (B) (Grant No. 26287087).

# Figure Captions

**FIG. 1 (Color online).** Density and polarization distributions characterized by linearly ((a1) and (a2)) and circularly polarized light ((b1) and (b2)) in the case of $E_g = 5\omega_0$ and $\Omega_{R0} = 2\omega_0$ for $t = 12\pi/\omega_0$. Current generations calculated by both distributions are plotted in FIG. (a3) and (b3), respectively. Red and green lines show temporal evolutions of the currents generated along the *x*- and *y*-axes.

**FIG. 2 (Color online).** Higher harmonic spectra generated from two-dimensional semiconductors: (a) multiphoton absorption regime, (b) ac Zener regime, and (c) semimetal regime. Red, green, and blue lines indicate the HHG spectra caused by the different elliptically polarized light.

**FIG. 3 (Color online).** Ellipticity dependences of higher-harmonic intensities focusing on (a) 7th harmonics in the multiphoton absorption regime, (b) 13th harmonics in the ac Zener regime, and (c) 28th harmonics in the semimetal regime. Red, green, and blue lines show the harmonics generated from the current components of $J_x(t)$, $J_y(t)$, and $J_{total}(t)$, respectively. These figures show that only parallel emissions with respect to the major axis exist in the multiphoton absorption and ac Zener regimes, while both parallel and perpendicular emissions exist in the semimetal regime.

**FIG. 4 (Color online).** Schematic diagrams of carrier trajectories caused by elliptically polarized electric fields in the cases of (a) ac Zener regime and (b) semimetal regime, respectively.

**FIG. 5 (Color online).** Upper left: Higher harmonic spectra generated from the current components of $J_x(t)$ (red lines), $J_y(t)$ (green lines), and $J_{total}(t)$ (blue lines) in the case of $\Omega_{R0} = 8\omega_0$ for $\epsilon = 0.3$. Upper right, lower left, and lower right: Harmonic intensities as a function of ellipticity focusing on 5th (a), 15th (b), and 30th harmonics (c). Red, green and blue lines show harmonics generated from the current components of $J_x(t)$, $J_y(t)$, and $J_{total}(t)$, respectively.

**FIG. 1**

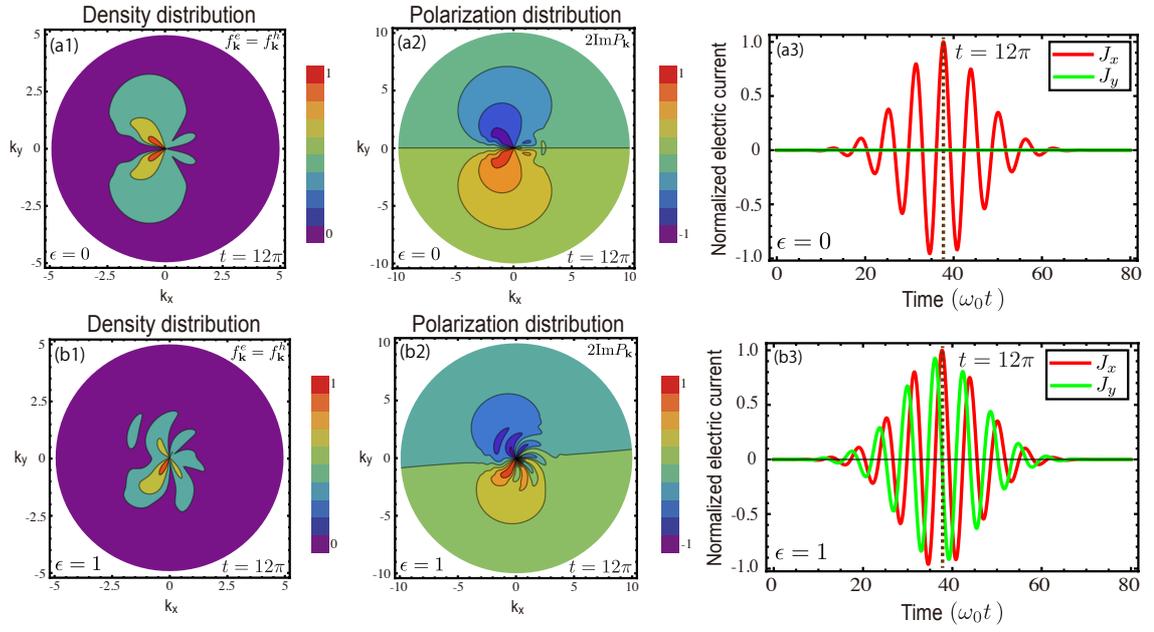

**FIG. 2**

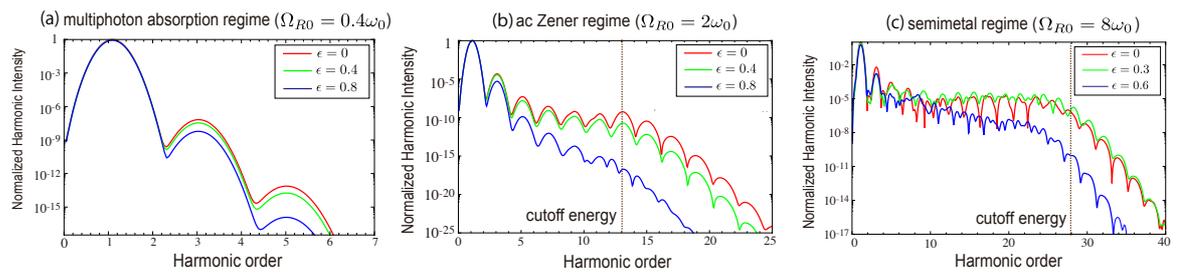

T. Tamaya *et al.*

**FIG. 3**

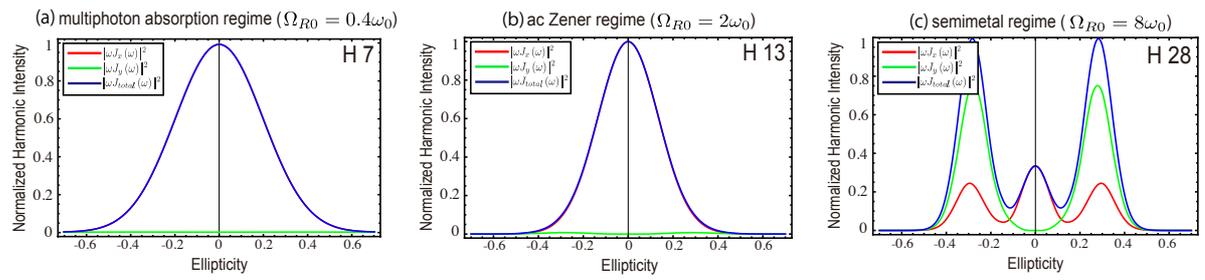

**FIG. 4**

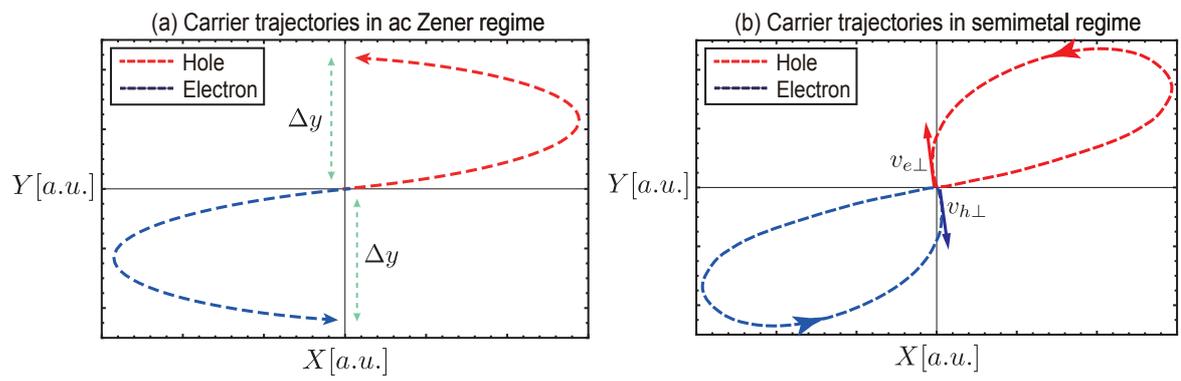

**FIG. 5**

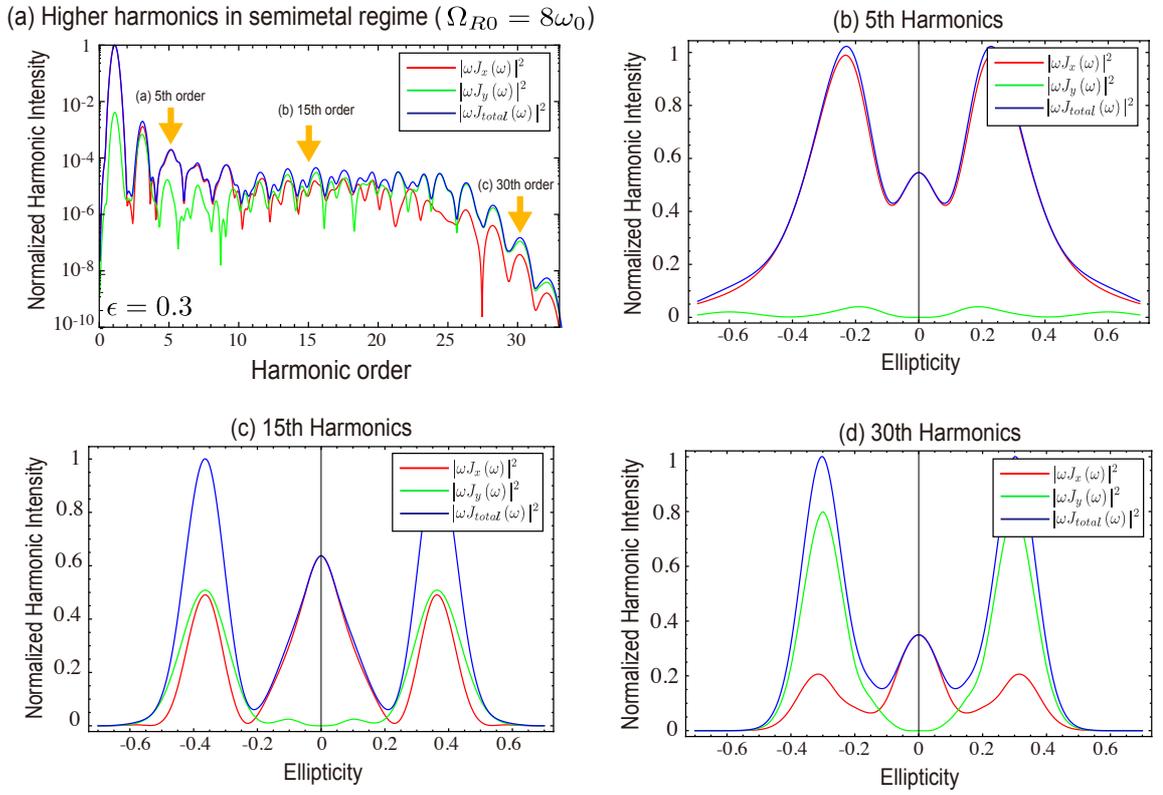